\documentclass[journal]{IEEEtran}

\usepackage{tikz}
\graphicspath{{images/}}
\usepackage{amsmath}
\usepackage{latexsym, amssymb}
\usepackage{graphicx}
\usepackage{amsmath, amsbsy}
\usepackage{amsopn, amstext}
\usepackage{ifpdf,hyperref}
\usepackage{cancel, color}
\usepackage{epstopdf}

\def\J{{\bf 1}}

\DeclareMathOperator{\rank}{rank}

\DeclareMathOperator{\Col}{Col}
\DeclareMathOperator{\Row}{Row}

\DeclareMathOperator{\lcm}{lcm}

\def\cal{\mathcal}

\def\ra{\rightarrow}

\def\a{\alpha}
\def\b{\beta}
\def\d{\delta}

\def\D{\Delta}

\def\0{{\bf 0}}

\newcommand{\R}{{\mathbb R}}

\def\dsum{\mathop{\sum}\limits}

\newtheorem{thm}{Theorem}[section]
\newtheorem{dfn}[thm]{Definition}
\newtheorem{prp}[thm]{Proposition}
\newtheorem{exa}[thm]{Example}
\newtheorem{lem}[thm]{Lemma}
\newtheorem{cor}[thm]{Corollary}
\newtheorem{rem}[thm]{Remark}
\newtheorem{alg}[thm]{Algorithm}

\begin{document}

\title{A Formula for Designing Zero-Determinant Strategies}

\author{Daizhan Cheng,
	\thanks{This work is supported partly by the National Natural Science Foundation of China (NSFC) under Grants 62073315, 61074114, and 61273013.}
	\thanks{Key Laboratory of Systems and Control, Academy of Mathematics and Systems Sciences, Chinese Academy of Sciences,
		Beijing 100190, P. R. China (e-mail: dcheng@iss.ac.cn).}
}
\maketitle

\begin{abstract}
A formula is presented for designing zero-determinant(ZD) strategies of general finite games,  which have $n$ players and players can have different numbers of strategies. To this end, using semi-tensor product (STP) of matrices, the profile evolutionary equation for repeated finite games is obtained. Starting from it, the ZD strategies are developed for general finite games, based on the same technique proposed by Press and Dyson \cite{pre12}. A formula is obtain to design ZD strategies for any player directly, ignoring the original ZD design process. Necessary and sufficient condition is obtained to ensure the effectiveness of the designed ZD strategies. As a consequence, it is also clear that  player $i$  is able to unilaterally design $|S_i|-1$ dominating linear relations about the expected payoffs of all players. Finally, the fictitious opponent player is proposed for networked evolutionary networks (NEGs). Using it, the ZD-strategies are applied to NEGs. It is surprising that an individual in a network may use ZD strategies to conflict the whole rest network.
\end{abstract}


%
\section{Introduction}

  Zero-determinant (ZD) strategy was firstly proposed by Press and Dyson \cite{pre12}, which shows that in an iterated prisoner's dilemma there exist strategies that dominate any evolutionary opponent. Since then it has attracted considerable attention from game theoretic community as well as computer, information, systems and control  communities. \cite{hao14} calls it ``an underway revolution in game theory", because it reveals that in a repeated game, a player can unilaterally control her opponent's payoff. The following up works include: (i) ZD strategy stability analysis \cite{szo14}; (ii) extension to multi-player and multi-strategy case \cite{hil14,he16}; (iii) games with continuous action spaces\cite{mca16}; (iv) application to public goods games \cite{pan15}, mining pole games \cite{cao19},  snowdrift game \cite{wan17}, etc.
A recent work \cite{tan21} tried to bypass working out determinant form and find linear restriction relations directly.

Though there are some efforts for extending ZD strategy technique, proposed by the pioneer work of Press and Dyson for prisoner's dilemma, to general cases \cite{hil14,he16}, to the author's best knowledge, a systematic description for most general games has not been obtained. The purpose of this paper is to present a general framework for general finite games.

Recently, a new matrix product, called the semi-tensor product (STP) of matrices, was proposed \cite{che11,che12}, and it has been applied to solve some problems in game theory, including the  modeling and analysis of networked evolutionary games (NEGs) \cite{che15}, providing a formula to verify whether a finite game is potential \cite{che14}, investigating the vector space structure of finite games and its orthogonal decompositions \cite{che17, hao18}, application to traffic congestion games \cite{zha21} and diffusion games \cite{li18}, etc., just to mention a few. Readers who are interested in the STP approach to finite games are referred to a survey paper \cite{chepr}.

 Using STP, this paper presents a profile evolutionary equation (PEE) for general finite games, which is essentially the same as the Markov matrix for the memory-one game in \cite{pre12}. Then a detailed design technique  and rigorous proofs are presented for this general case, which are  generalizations  of those proposed firstly by \cite{pre12}. Necessary and sufficient conditions are provided for the effectiveness of ZD strategies. As a by product, we also prove that if a player has $k_i$ strategies she can provide unilaterally $k_i-1$ linear payoff relations using ZD strategies.

The main contribution of  this paper is the formula, which  provides an algebraic equation for designing ZD strategies, which does not involve any original determinant.

Finally, the ZD strategies for Networked evolutionary games (NEGs) are investigated. By proposing and using fictitious opponent player (FOP) the networked evolutionary games can be formulated as a two player game, where a player, say, player $i$,  plays fights with the FOP, who represents the whole network except player $i$.  The ZD strategies for player $i$ ar designed for $i$ with FOP.

The rest of this paper is organized as follows: A brief survey is given in Section 2. Then It is used to develop profile evolutionary equation (PEE) of finite evolutionary games. Finally, some properties of transition matrix of PPE are investigated, which are important for designing ZD strategies. Section 3 deduces a general formula for designing ZD strategies. Necessary and sufficient conditions for the designed ZD strategies to be effective is presented. Some numerical examples are followed to describe the design procedure. The fictitious opponent player (FOP) is proposed in Section 4 for networked evolutionary games (NEGs). Using FOP, the technique of ZD strategies has been applied to NEGs.
Section 5 is a brief conclusion.

Before ending this section, a list of notations is presented.

\begin{itemize}

\item ${\cal M}_{m\times n}$: set of $m\times n$ dimensional real matrices.

\item $M^*$: the adjoint matrix of $M$.

\item $\sigma(M)$: the set of eigenvalues of $M$.

\item $\rho(M)$: the spectral radius of $M$.

\item $M>0$ ($M\geq 0$): all entries of $M$ are positive (non-negative), i.e., $M_{i,j}>0$ ($M_{i,j}\geq 0$), $\forall i,j$.

\item $\ltimes$: STP of matrices.

\item $\Col(A)$ ($\Row(A)$) : the set of columns (rows) of ~$A$; $\Col_i(A)$ ($\Row_i(A)$): the $i$-th column (row) of ~$A$.

\item ${\cal D}_k$: ${\cal D}_k=\{1,2,\cdots,k\}$.
\item
$\d_k^i$: The $i$-th column of identity matrix $I_k$. $\d_k^0$ is for a zero vector of dimension $k$.

\item
$\D_k$: $\D_k=\Col(I_k)=\left\{\d_k^i\;|\;i=1,\cdots,k\right\}$

\item $L\in {\cal M}_{m\times n}$ is called a logical matrix, if $\Col(L)\subset \D_m$.

\item Let $L=\left[\d_m^{i_1},\d_m^{i_2},\cdots,\d_m^{i_n}\right]$, it is briefly denoted by
$L=\d_m[i_1,i_2,\cdots,i_n]$.

\item $\Upsilon^m$:  set of $m$ dimensional (column) random vectors.

\item $\Upsilon_{n\times r}$: set of $n\times n$ (column) random matrices.

\item
${\cal G}_{[n;k_1,k_2,\cdots,k_n]}$: set of finite non-cooperative games with $n$ players, and player $i$ has $k_i$ strategies, $i=1,2,\cdots,n$.

\vskip 2mm

\end{itemize}

\section{Modeling of Finite  Evolutionary Games}

\subsection{A Brief Survey on STP}

\begin{dfn}\label{d2.1.1}\cite{che11,che12}
Let $M\in {\cal M}_{m\times n}$, $N\in {\cal M}_{p\times q}$, and
$t:=\lcm(n,p)$ be the least common multiple of $n$ and $p$.
Then  the STP   of $M$ and $N$ is defined as
\begin{align}\label{2.1.1}
M\ltimes N:=\left( M\otimes I_{t/n}\right)\left(N\otimes I_{t/p}\right)\in {\cal M}_{(mt/n) \times (qt/p)}.
\end{align}
\end{dfn}

\begin{rem}\label{r2.1.2}
\begin{itemize}
\item[(i)] STP is a generalization of conventional matrix product. That is, if $n=p$, then $M\ltimes N=MN$. Hence in most cases the symbol $\ltimes$ can be omitted. It is applicable to two arbitrary matrices.

\item[(ii)] As a generalization, STP keeps all major properties of conventional matrix product available, including  associativity, distributivity, etc.

\item[(iii)] STP is a fundamental tool in this approach, and all the matrix products used in this paper without product symbol are assumed to be STP. Hence, in this paper the product of two arbitrary size matrices is well proposed.
\end{itemize}
\end{rem}

Let $f:{\cal D}_m\ra {\cal D}_n$ be a mapping from a finite set to another finite set. Then we can identify $j\in {\cal D}_m$ with its vector form $\vec{j}:=\d_{m}^j\in \D_{m}$. In this way, $f$ can be considered as a mapping $f: \D_m\ra \D_n$. In the sequel $\vec{j}$ is simply denoted by $j$ again if there is no possible confusion.

\begin{prp}\label{p2.1.3} Let $f:{\cal D}_m\ra {\cal D}_n$. Then there exists a unique matrix $M_f\in {\cal L}_{m\times n}$, called the structure matrix of $f$, such that as the arguments are expressed into their vector forms, we have
\begin{align}\label{2.1.2}
f(x)=M_fx.
\end{align}
\end{prp}

As a corollary,  Proposition \ref{p2.1.3} can be extended into more general form.

\begin{cor}\label{c1.2.4} Let $x_i\in {\cal D}_{k_i}$, $i=1,2,\cdots,n$, $y_j\in {\cal D}_{p_j}$, $j=1,2,\cdots,m$, and $x=\ltimes_{i=1}^nx_i$, $y=\ltimes_{j=1}^my_j$. Assume
$$
y_j=f_j(x_1,x_2,\cdots,x_n),\quad j=1,2,\cdots,m,
$$
which have their vector forms as
\begin{align}\label{2.1.4}
y_j=M_j\ltimes_{i=1}^nx_i, \quad j=1,2,\cdots,m.
\end{align}
Then there exists a unique matrix $M_F$, called the structure matrix of the mapping $F=(f_1,f_2,\cdots,f_m)$,
such that
\begin{align}\label{2.1.5}
y=M_Fx,
\end{align}
where
$$
M_F=M_1*M_2*\cdots*M_n \in {\cal L}_{\rho\times \kappa},
$$
and $\rho=\prod_{j=1}^mp_j$, $\kappa=\prod_{i=1}^nk_i$,
and  $*$ is Kratri-Rao product of matrices.
\footnote{Let $A\in {\cal M}_{s\times n}$, $B\in {\cal M}_{t\times n}$. Then the Khatri-Rao product of $A$ and $B$, denoted by $A*B$, is defined by \cite{che12}
$$
\Col_i(A*B)=\Col_i(A)\Col_i(B),\quad i=1,2,\cdots,n.
$$
}

\end{cor}

Similarly, we have the following result:

\begin{cor}\label{c1.2.5} Let $x_i\in \Upsilon_{k_i}$, $i=1,2,\cdots,n$ and $y_j\in \Upsilon_{p_j}$, $j=1,2,\cdots,m$, and
\begin{align}\label{2.1.6}
y_j=M_jx, \quad j=1,2,\cdots,m,
\end{align}
where $M_j\in \Upsilon_{\kappa\times p_j}$.
Then there exists a unique matrix $M_F$ such that
\begin{align}\label{2.1.7}
y=M_Fx,
\end{align}
where
$$
M_F=M_1*M_2*\cdots*M_n\in \Upsilon_{\rho\times \kappa},
$$
which is also called the structure matrix of the mapping $F=(f_1,f_2,\cdots,f_m)$.
\end{cor}

\subsection{PEE of Finite Games}

\begin{dfn}\label{d2.2.1}
Consider a finite game $G=(N, S,C)$, where
\begin{itemize}
\item[(i)] $N=\{1,2,\cdots,n\}$ is the set of players.

\item [(ii)] $S=\prod_{i=1}^n S_i$ is the profile, where
$$
S_i=\{1,2,\cdots,k_i\},\quad i=1,2,\cdots,n,
$$
is the strategies (or actions) of player $i$.

\item[(iii)] $C=(c_1,c_2,\cdots,c_n)$, where
$$
c_i:S\ra \R,\quad i=1,2,\cdots,n,
$$
is the payoff (or utility, cost) function of player $i$.
\end{itemize}
\end{dfn}

The set of such games is described by ${\cal G}_{[n;k_1,k_2,\cdots,k_n]}$.

A matrix formulation of the evolution game  $G\in {\cal G}_{[n;k_1,k_2,\cdots,k_n]}$ is described as follows \cite{che15}:

\begin{itemize}
\item[(i)] Identifying $j\in S_i$ with $\d_{k_i}^j\in \D_{k_i}$, then $S_i\sim \D_{k_i}$.

\item[(ii)] Setting $\kappa=\prod_{i=1}^nk_i$, then $S\sim \D_{\kappa}=\prod_{i=1}^n \D_{k_i}$.

\item[(iii)] Let $x_i\in \D_{k_i}$ be the vector form of a strategy for player $i$. Then $x=\ltimes_{i=1}^nx_i\in \D_{\kappa}$ is a profile.

\item[(iv)] For each player's payoff function $c_i$, there exists a unique row vector $V^c_i\in \R^{\kappa}$ such that
\begin{align}\label{2.2.1}
c_i(x)=V^c_ix,\quad i=1,2,\cdots,n.
\end{align}
\end{itemize}

Now consider an evolutionary game $G^e$ of $G$, which stands for (infinitely) repeated $G$.  Then each player can determine his action at $t+1$ using historic knowledge. It was proved in \cite{pre12} (see also \cite{hao14}) that:``the shortest memory player sets the rule of the game, which means the long-memory strategies have no advantages over the memory-one strategies". Based on this observation, the strategy updating rule is assumed Markov-like. That is, the strategy of player $i$ at time $t+1$ depends on the profile at $t$ only. Then we have \cite{che15}
\begin{align}\label{2.2.2}
x_i(t+1)=L_ix(t),\quad i=1,2,\cdots,n.
\end{align}

Two types of strategies are commonly used:
\begin{itemize}
\item Pure strategy:
$$
L_i\in {\cal L}_{k_i\times \kappa},\quad i=1,2,\cdots,n.
$$
\item Mixed Strategy:
$$
L_i\in \Upsilon_{k_i\times \kappa},\quad i=1,2,\cdots,n.
$$
\end{itemize}

Multiplying (by STP) all equations in (\ref{2.2.2}) together yields
\begin{align}\label{2.2.3}
x(t+1)=Lx(t),
\end{align}
where
$$
L=L_1*L_2*\cdots*L_n.
$$

In pure strategy case $L\in {\cal L}_{\kappa\times \kappa}$ and in mixed strategy case $L\in \Upsilon_{\kappa\times \kappa}$.

 In mixed strategy case $x(t)$ can be considered as a distribution of profiles at time $t$. If we take into consideration that $\d_{\kappa}^i$ is used to express the $i$-th profile, then $x(t)$ can also be considered as the expected value of profile at time $t$.

In this paper we concern only mixed strategy case. Now what a player can manipulate is his own strategy updating rule. That is, player $i$ can only choose his $L_i$.

We arrange profiles in an alphabetic order as
$$
\begin{array}{ccl}
S&=&\{(s_1,s_2,\cdots,s_n)\;|\;s_i\in S_i,~i=1,2,\cdots,n\}\\
~&=&\{(1,1,\cdots,1),(1,1,\cdots,2),\cdots,(k_1,k_2,\cdots,k_n)\}\\
~&:=&\{s^1,s^2,\cdots,s^{\kappa}\}.
\end{array}
$$

Denote the probability of player $i$ choosing strategy $j$ at time $t+1$ under the situation that the profile at time $t$ is $s^r$ as
\begin{align}\label{2.2.4}
p^r_{i,j}=Prob(x_i(t+1)=j\;|\;x(t)=s^r).
\end{align}
Then we have the strategy evolutionary equation (SEE) of player $i$ as
\begin{align}\label{2.2.5}
x_i(t+1)=L_ix(t),
\end{align}
where
\begin{align}\label{2.2.6}
\begin{array}{l}
L_i=\begin{bmatrix}
p^1_{i,1}&p^2_{i,1}&\cdots&p^{\kappa}_{i,1}\\
p^1_{i,2}&p^2_{i,2}&\cdots&p^{\kappa}_{i,2}\\
\vdots&~&~&~\\
p^1_{i,k_i}&p^2_{i,k_i}&\cdots&p^{\kappa}_{i,k_i}\\
\end{bmatrix}\\
~~~~~~~~~~~~i=1,2,\cdots,n.
\end{array}
\end{align}

In fact, in $G^e$ we assume the strategies of player $i$ are
$$
S^e_i=\{s^r\ra j\;|\;s^r\in S,r=1,2,\cdots,\kappa, j=1,2,\cdots,k_i\},
$$
where $s^r\ra j$ means where $x(t)=s^r$ then $x_i(t+1)=j\in S_i$.

 $L_i$ comes from the definition  (\ref{2.2.4})  immediately.

According to Corollary \ref{c1.2.5}, we have PEE as
\begin{align}\label{2.2.7}
x(t+1)=Lx(t),
\end{align}
where the transition matrix
\begin{align}\label{2.2.701}
L=L_1*L_2*\cdots*L_n.
\end{align}

We give a simple example to calculate $L$.

\begin{exa}\label{e1.2.6} Consider the repeated prisoners' dilemma. Let $p^r_{i,j}$ be the probability of player $i$ taking strategy $j\in \{C,D\}\sim \{1,2\}$ under the condition $s^r$. Then a straightforward computation shows that
$$
\begin{array}{l}
x_1(t+1)=L_1x(t),\\
x_2(t+1)=L_2x(t),
\end{array}
$$
where
$$
L_1=\begin{bmatrix}
p^1_{1,1}&p^2_{1,1}\\
p^1_{1,2}&p^2_{1,2}\\
\end{bmatrix},
$$
$$
L_2=\begin{bmatrix}
p^1_{2,1}&p^2_{2,1}\\
p^1_{2,2}&p^2_{2,2}\\
\end{bmatrix}.
$$
Denote by
$$
\begin{array}{l}
p_i=p^i_{1,1},\\
q_i=p^i_{2,1},\quad i=1,2,3,4.
\end{array}
$$
It follows that
$$
\begin{array}{l}
p^i_{1,2}=1-p^i_{1,1}=1-p_i,\\
p^i_{2,2}=1-p^i_{2,1}=1-q_i,\quad i=1,2,3,4.
\end{array}
$$
Then we have
\begin{align}\label{2.2.8}
\begin{array}{l}
L=L_1*L_2\\
~=\begin{bmatrix}
p^1_{1,1}p^1_{2,1}&p^2_{1,1}p^2_{2,1}&p^3_{1,1}p^3_{2,1}&p^4_{1,1}p^4_{2,1}\\
p^1_{1,1}p^1_{2,2}&p^2_{1,1}p^2_{2,2}&p^3_{1,1}p^3_{2,2}&p^4_{1,1}p^4_{2,2}\\
p^1_{1,2}p^1_{2,1}&p^2_{1,2}p^2_{2,1}&p^3_{1,2}p^3_{2,1}&p^4_{1,2}p^4_{2,1}\\
p^1_{1,2}p^1_{2,2}&p^2_{1,2}p^2_{2,2}&p^3_{1,2}p^3_{2,2}&p^4_{1,2}p^1_{2,2}\\
\end{bmatrix}=\\
\begin{tiny}
\begin{bmatrix}
p_1q_1&p_1q_2&p_2q_1&p_2q_2\\
p_1(1-q_1)&p_1(1-q_2)&p_2(1-q_1)&p_2(1-q_2)\\
(1-p_1)q_1&(1-p_1)q_2&(1-p_2)q_1&(1-p_2)q_2\\
(1-p_1)(1-q_1)&(1-p_1)(1-q_2)&(1-p_2)(1-q_1)&(1-p_2)(1-q_2)\\
\end{bmatrix}
\end{tiny}
\end{array}
\end{align}
\end{exa}

\begin{rem}\label{r1.2.7}
It is easy to verify that the transition matrix in PEE (refer to (\ref{2.2.8})) is essentially the transpose of the Markov matrix for the memory one game in \cite{pre12}. (Corresponding to the ``column order" of \cite{pre12} there is a ``row order" change in (\ref{2.2.8}). This is because our profiles are ordered in alphabetic as $CC,~CD, ~DC, ~DD$, while \cite{pre12} uses the order $CC,~DC,~CD,~DD$.
\end{rem}

\subsection{Properties of PEE}

In this subsection we investigate some properties of the transition matrix $L$ of PEE, which are required for designing ZD strategies. As we discussed in Remark \ref{r1.2.7}, $L$ is the same as the Markov matrix for the memory one game in \cite{pre12}, (only with a transpose). So $L$ is a column random matrix. This difference does not affect the following discussion. Hence the following argument is a mimic of the corresponding argument in  \cite{pre12}. What we are going to do is to extend it to general case and put it on a solid mathematical foundation.

In the sequel, we need an assumption on $L$. To present it, some preparation is necessary.

A random square matrix $M$ is called a primitive matrix if there exists a finite integer $s>0$ such that $M^s>0$ \cite{hor86}.

Some nice properties of primitive matrix are cited as follows:

\begin{prp}\label{p2.3.1}\cite{hor86} Let $L$ be a primitive stochastic matrix. Then
\begin{itemize}
\item[(i)]  $\rho(P)=1$ and there exists a unique $\lambda\in \sigma(P)$ such that $|\lambda|=1$.

\item[(ii)]
\begin{align}\label{2.3.1}
\lim_{t\ra \infty}L^t=P>0.
\end{align}
Moreover, $P=uv^T$, where $Lu=u$, $u>0$, $L^Tv=v$, $v>0$.
\end{itemize}
\end{prp}

We are ready to present our fundamental assumption.

\vskip 2mm

{\bf Assumption A-1:} $L$ is primitive.

\vskip 2mm

\begin{rem}\label{r2.3.2}
\begin{itemize}
\item[(i)] A-1 is not always true. For instance, consider (\ref{2.2.8}) and let $p^2_{1,1}=0$, $p^3_{1,1}=0$, $p^4_{1,1}=0$, and $p^4_{2,1}=0$. Then $L$ is not primitive.
\item[(ii)] If $0<p^r_{i,j}<1$, $\forall r,i,j$, then a straightforward verification shows tat $L$  is primitive. So A-1 is always true except a zero-measure set.
\item[(iii)] According to Proposition \ref{p2.3.1}, we have the following immediate conclusions: (a), if $L$ is primitive, then
\begin{align}\label{2.3.2}
\rank(L-I_{\kappa})=\kappa-1.
\end{align}
(b) There exists $P=uv^T$, where $Lu=u$, $u>0$, $L^Tv=v$, $v>0$, such that (\ref{2.3.1}) holds. That is,
\begin{align}\label{2.3.3}
\lim_{t\ra \infty}L^t=uv^T.
\end{align}

\end{itemize}
\end{rem}

\begin{prp}\label{p2.3.3} Let $L$ be an $\kappa\times \kappa$ column primitive stochastic matrix.
Define $M:=L-I_{\kappa}$ and denote by $M^*$  its adjoint matrix. Then
\begin{itemize}
\item[(i)]  $\rank(M^*)=1$.
\item[(ii)]
\begin{align}\label{2.3.4}
\Col_j(M^*)\neq 0,\quad j=1,2,\cdots,\kappa.
\end{align}
\end{itemize}
\end{prp}

The proof and all other proofs can be found in Appendix.

\begin{prp}\label{p2.3.4} Consider the profile evolutionary equation (\ref{2.2.7}). If $L$ is primitive, then
\begin{align}\label{2.3.5}
x^*:=\lim_{t\ra \infty}x(t)=u/\|u\|,
\end{align}
where $u$ comes from (\ref{2.3.3}).
\end{prp}

Hereafter, we assume $u$ has been normalized. Then $x^*=u$ is the only normalized eigenvector of $L$ corresponding to eigenvalue $1$.

\begin{prp}\label{p2.3.5} Assume $L$ is primitive, then
\begin{align}\label{2.3.6}
\Col_j(M^*)\propto u,\quad \forall j.
\end{align}
\end{prp}

Combining (\ref{2.3.4}) and (\ref{2.3.6}) yields
\begin{align}\label{2.3.7}
\Col_j(M^*)=\mu_ju,\quad \mu_j\neq 0,\;\; j=1,2,\cdots,\kappa.
\end{align}

\section{Design of ZD Strategies for Evolutionary Games}

\subsection{A Universal Formula for ZD-strategies}

Consider the transition matrix $L$ of PEE (\ref{2.2.7}). For statement ease, we define two sets of parameters as
\begin{align}\label{3.1}
\kappa_{(i)}:=
\begin{cases}
1,\quad i=1,\\
\prod_{j=1}^{i-1}k_j,\quad i=2,3,\cdots,n.
\end{cases}
\end{align}
\begin{align}\label{3.2}
\kappa^{(i)}:=
\begin{cases}
0,\quad i=0,\\
\prod_{j=i+1}^{n}k_j,\quad i=1,2,\cdots,n-1,\\
1,\quad i=n.
\end{cases}
\end{align}

Using them, for each pair $i,j$, $i=1,\cdots,n$, $j=1,2,\cdots,k_i$,
we define an index set as:
\begin{align}\label{3.3}
\begin{array}{ccl}
\Phi_{i,j}&:=&\left\{s=(\a-1)\kappa^{(i-1)}+(j-1)\kappa^{(i)}+\b \right.\;|\;\\
~&~&\left.\a=1,2,\cdots,\kappa_{(i)};
~\b=1,2,\cdots,\kappa^{(i)}\right\}.
\end{array}
\end{align}

Then we have the following result:

\begin{lem}\label{l3.1} Let $M=L-I_{\kappa}$. Define
\begin{align}\label{3.4}
\begin{array}{ccl}
\Xi_{i,j}&:=&\dsum_{s\in \Phi_{i,j}}\Row_{s}(M),\\
~&~&~~~~~~~~~~~i=1,2,\cdots,n;~j=1,2,\cdots,k_i.
\end{array}
\end{align}
Then
\begin{align}\label{3.5}
\Xi_{i,j}=p_{i,j}-\xi_{i,j},\quad, i=1,2,\cdots, n;\;j=1,2,\cdots,k_i,
\end{align}
where
$$
p_{i,j}=(p^1_{i,j},p^2_{i,j},\cdots,p^{\kappa}_{i,j}),
$$
and $\xi_{i,j}\in {\cal B}^{\kappa}$ is defined as follows:
\begin{align}\label{3.6}
\xi_{i,j}(s)=
\begin{cases}
1,\quad s\in \Phi_{i,j},\\
0,\quad s\not\in \Phi_{i,j}.
\end{cases}
\end{align}
\end{lem}

It is obvious from  Lemma \ref{l3.1} that $\Xi_{i,j}$ can be used to design $p_{i,j}$ for player $i$. A general design formula is presented in the following algorithm:

\begin{alg}\label{a3.2} Consider an evolutionary game $G^e$, where  $G\in {\cal G}_{[n;k_1,k_2,\cdots,k_n]}$. Assume player $i$ is aimed at a set of linear relations on the expected payoffs as
\begin{align}\label{3.7}
\begin{array}{l}
\ell_{i,j}(Ec_1,Ec_2,\cdots,Ec_n,1),\\
\quad 1\leq i\leq n; j=1,2,\cdots,k_i-1,
\end{array}
\end{align}
where $\ell_{i,j}$ is a linear equation. Then his ZD strategies can be designed as
\begin{align}\label{3.8}
\begin{array}{ccl}
p_{i,j}&=&(p^1_{i,j},p^2_{i,j},\cdots,p^{\kappa}_{i,j})\\
~&=&\mu_{i,j}\ell_{i,j}\left(V^c_1,V^c_2,\cdots,V^c_n,\J_{\kappa}^T\right)+\xi_{i,j},\\
~&~& \quad j=1,2,\cdots,k_i-1,
\end{array}
\end{align}
where $\mu_{i,j}\neq 0$ are adjustable parameters.
\end{alg}

Equation (\ref{3.8}) is a fundamental formula, which provides a convenient way to design ZD strategies for our preassigned purposes.

\begin{dfn}\label{d3.8} A set of ZD strategies is rational, if  the following rationality conditions are satisfied.
\begin{itemize}
\item[(i)]
\begin{align}\label{3.9}
0\leq p_{i,j}\leq 1,\quad j=1,2,\cdots,k_i-1.
\end{align}
\item[(ii)]
\begin{align}\label{3.10}
0\leq \dsum_{j=1}^{k_i-1}p_{i,j}\leq 1.
\end{align}
\end{itemize}
\end{dfn}

\begin{rem}\label{r3.9}
\begin{itemize}
\item[(i)] It is obvious that rationality is a fundamental requirement. Non-rational strategies are meaningless.

\item[(ii)] It is clear that player $i$ can unilaterally design at most $|S_i|-1$ linear relations. Because when
$p_{i,j}$, $j<|S_i|$ are all determined, $p_{i,|S_i|}$ is uniquely determined by
$$
p_{i,|S_i|}=\J^T_{\kappa}-\dsum_{j=1}^{|S_i|-1}p_{i,j}.
$$

\item[(iii)] Of course, player $i$ need not to design $|S_i|-1$ relations. Is he intends to design $r<|S_i|-1$ relations, equation (\ref{3.10}) has to be modified by reducing the summation to $r$ items.

\item[(iv)] The ZD design formula (\ref{3.8}) can be used simultaneously by multi-players, or even all $n$ players.
\end{itemize}
\end{rem}

Even though a set of ZD strategies is rational, it may not be effective. That is, the goal (\ref{3.7}) may not be reached. We need the following result:

\begin{thm}\label{t3.10} Consider an evolutionary game, $G^e$, where $G\in {\cal G}_{[n;k_1,k_2,\cdots,k_n]}$.  A set of ZD strategies designed by formula (\ref{3.8}) is effective, if and only if,
\begin{itemize}
\item[(i)] There exists an $u\in \Upsilon^{\kappa}$ such that
\begin{align}\label{3.11}
\lim_{t\ra\infty}L^t=u \J^T_{\kappa}.
\end{align}
\item[(ii)]
\begin{align}\label{3.12}
\rank(L-I_{\kappa})=\kappa-1.
\end{align}
\end{itemize}
\end{thm}

Note that verifying the two conditions in Theorem \ref{t3.10} is not an easy job. Hence we may replace them by the following one.

\begin{cor}\label{c3.11} Consider  an evolutionary game, $G^e$, where $G\in {\cal G}_{[n;k_1,k_2,\cdots,k_n]}$. If $L$ is  primitive, then the set of ZD strategies designed by formula (\ref{3.8}) is effective.
\end{cor}

\begin{rem}\label{r3.12}
\begin{itemize}
\item[(i)] Even though  primitivity of $L$ is only a sufficient condition, it is almost necessity because only a zero-measure set of $L$ may not be primitive. That is, all non-primitive $L$ have certain boundary values as $p^r_{i,j}\in \{0,1\}$ for some $(r,i,j)$. So the designer, who intends to use ZD strategies, is better to avoid using such values.

\item[(ii)] Any player can not unilaterally make the conditions in Theorem \ref{t3.10} satisfied. It depends on other players' strategies. What the player $i$ can do is to do his best, that is, to ensure his designed rows, $\Xi_{i,j}$, $j=1,2,\cdots,k_i$ are linearly independent. (A Chinese idiom says that ``Mou~Shi~Zai~Ren, Cheng~Shi~Zai~Tian", (Man proposes, God disposes. That is the situation for ZD strategy Designer.)).
\end{itemize}
\end{rem}

\subsection{Numerical Examples}

In the following, we discuss some numerical examples:
\begin{exa}\label{e4.1}

Consider a $G\in {\cal G}_{[3;2,3,2]}$. Since
$k_1=2$, $k_2=3$, and $k_3=2$, using (\ref{3.1}) and (\ref{3.2}), it is easy to calculate that
\begin{align}\label{4.1}
\begin{array}{l}
\kappa_{(1)}=1,\quad \kappa_{(2)}=2,\quad \kappa_{(3)}=6;\\
\kappa^{(0)}=0,\kappa^{(1)}=6,\quad \kappa^{(2)}=2,\quad \kappa^{(3)}=1.\\
\end{array}
\end{align}
According to (\ref{3.3}), we have
\begin{align}\label{4.2}
\begin{array}{l}
\Phi_{1,1}=\{1,2,3,4,5,6\},\\
\Phi_{1,2}=\{7,8,9,10,11,12\},\\
\Phi_{2,1}=\{1,2,7,8\},\\
\Phi_{2,2}=\{3,4,9,10\},\\
\Phi_{2,3}=\{5,6,11,12\},\\
\Phi_{3,1}=\{1,3,5,7,9,11\},\\
\Phi_{3,2}=\{2,4,6,8,10,12\}.
\end{array}
\end{align}
\begin{align}\label{4.3}
\begin{array}{l}
\xi_{1,1}=[1,1,1,1,1,1,0,0,0,0,0,0],\\
\xi_{1,2}=[0,0,0,0,0,0,1,1,1,1,1,1],\\
\xi_{2,1}=[1,1,0,0,0,0,1,1,0,0,0,0],\\
\xi_{2,2}=[0,0,1,1,0,0,0,0,1,1,0,0],\\
\xi_{2,3}=[0,0,0,0,1,1,0,0,0,0,1,1],\\
\xi_{3,1}=[1,0,1,0,1,0,1,0,1,0,1,0],\\
\xi_{3,2}=[0,1,0,1,0,1,0,1,0,1,0,1].\\
\end{array}
\end{align}

These parameters depend on the type of games, precisely speaking, they depend on $\{n;k_1,k_2,\cdots,k_n\}$ only. They are  independent on particular games.

\begin{itemize}
\item Pinning Strategy:

Assume the payoff vectors are:
$$
\begin{array}{ccl}
V^c_1&=&[-3,-0.5,6,9,8,7,-4,-4.5,5,6.5,5,7],\\
V^c_2&=&[4,-1,-5,7.5,2,3.5,8,-4,5,8,9,-2],\\
V^c_3&=&[9,5,-6,-5.5,5.5,8,8.5,5.5,-0,-3.5,\\
~&~&~4.5,7].
\end{array}
$$

Set
\begin{align}\label{4.4}
\begin{array}{ccl}
p_{2,1}&:=&(0.1)*V^c_1-(0.4)*\J_{12}^T+\xi_{2,1}\\
~&=&[0.3,0.55,0.2,0.5, 0.4, 0.3, 0.2, 0.15, 0.1,\\
~&~&~0.25,0.1, 0.3]\\
p_{2,2}&=&(0.1)*V^c_3+(0.3)*\J_{12}^T+\xi_{2,2}\\
~&=&[0.6, 0.2, 0.1, 0.15, 0.25, 0.5, 0.55, 0.25,\\
~&~&~ 0.7, 0.35, 0.15, 0.4].
\end{array}
\end{align}

Then it is ready to verify that the ZD strategies designed in (\ref{4.4}) are rational.

\item[(ii)] Extortion Strategy:

Consider a $G\in {\cal G}_{[3;2,3,2]}$ again.  Assume the payoff structure vectors are as follows:

$$
\begin{array}{ccl}
V^c_1&=&[16, 11, -4, -8, -2, -10.3,11.4,18.5,\\
~&~&~ 1.2,-3,-2.5,1.5],\\
V^c_2&=&[3, 2, -1, 0, 5, -6, 4, 3, 3, 1, -1, 7],\\
V^c_3&=&[-2.9,0,6.8,7.1,2,-9.4,-8.2,0.4,\\
~&~&~4.6,6.1,-2,2.3].
\end{array}
$$

Assume player 2 want to design an extortion strategy against both players 1 and 3. He may design
$$
\begin{array}{l}
Ec_2-r=\kappa_1(Ec_1-r),\\
Ec_2-r=\kappa_2(Ec_3-r).
\end{array}
$$

To this end, he needs to design
$$
\begin{array}{l}
p_{2,1}-\xi_{2,1}=\mu_1\left[(V^c_2-r\J_{12}^T)-\kappa_1(V^c_1-r\J_{12}^T)\right],\\
p_{2,3}-\xi_{2,3}=\mu_2\left[(V^c_2-r\J_{12}^T)-\kappa_2(V^c_3-r\J_{12}^T)\right].\\
\end{array}
$$

Choosing $\mu_1=0.05$, $\mu_2=0.1$, $r=1$, $\kappa_1=1.1$, $\kappa_2=1.2$,
it follows that

\begin{align}\label{4.5}
\begin{array}{ccl}
p_{2,1}&=&(0.275,0.5,0.175,0.445,0.365,0.2715,\\
~&~&~0.178,0.1375,0.0890,0.22,0.0925,0.2725],\\
p_{2,2}&=&(0.668,0.22,0.104,0.168,0.28,0.548,\\
~&~&0.604,0.272,0.768,0.388,0.16,0.444].\\
\end{array}
\end{align}

The ZD strategies designed in (\ref{4.5}) are also rational.

\end{itemize}

\end{exa}

\begin{rem}\label{r4.2}
\begin{itemize}
\item[(i)] In general, to design a set of rational ZD strategies is not an easy job. To determine related parameters we need to solve a set of linear inequalities.

\item[(ii)] To verify Lemma \ref{l3.1}, we calculate the  matrix $M=L-I_{\kappa}$ for Example \ref{e4.1} as follows:
\begin{tiny}
\begin{align}\label{4.6}
M=\left[
\begin{array}{llll}
p^1_{1,1}p^1_{2,1}p^1_{3,1}-1&p^2_{1,1}p^2_{2,1}p^2_{3,1}&\cdots&p^{12}_{1,1}p^{12}_{2,1}p^{12}_{3,1}\\
p^1_{1,1}p^1_{2,1}p^1_{3,2}&p^2_{1,1}p^2_{2,1}p^2_{3,2}-1&\cdots&p^{12}_{1,1}p^{12}_{2,1}p^{12}_{3,2}\\
p^1_{1,1}p^1_{2,2}p^1_{3,1}&p^2_{1,1}p^2_{2,2}p^2_{3,1}&\cdots&p^{12}_{1,1}p^{12}_{2,2}p^{12}_{3,1}\\
p^1_{1,1}p^1_{2,2}p^1_{3,2}&p^2_{1,1}p^2_{2,2}p^2_{3,2}&\cdots&p^{12}_{1,1}p^{12}_{2,2}p^{12}_{3,2}\\
p^1_{1,1}p^1_{2,3}p^1_{3,1}&p^2_{1,1}p^2_{2,3}p^2_{3,1}&\cdots&p^{12}_{1,1}p^{12}_{2,3}p^{12}_{3,1}\\
p^1_{1,1}p^1_{2,3}p^1_{3,2}&p^2_{1,1}p^2_{2,1}p^2_{3,2}&\cdots&p^{12}_{1,1}p^{12}_{2,1}p^{12}_{3,2}\\
p^1_{1,2}p^1_{2,1}p^1_{3,1}&p^2_{1,2}p^2_{2,1}p^2_{3,1}&\cdots&p^{12}_{1,2}p^{12}_{2,1}p^{12}_{3,1}\\
p^1_{1,2}p^1_{2,1}p^1_{3,2}&p^2_{1,2}p^2_{2,1}p^2_{3,2}&\cdots&p^{12}_{1,2}p^{12}_{2,1}p^{12}_{3,2}\\
p^1_{1,2}p^1_{2,2}p^1_{3,1}&p^2_{1,2}p^2_{2,2}p^2_{3,1}&\cdots&p^{12}_{1,2}p^{12}_{2,2}p^{12}_{3,1}\\
p^1_{1,2}p^1_{2,2}p^1_{3,2}&p^2_{1,2}p^2_{2,2}p^2_{3,2}&\cdots&p^{12}_{1,2}p^{12}_{2,2}p^{12}_{3,2}\\
p^1_{1,2}p^1_{2,3}p^1_{3,1}&p^2_{1,2}p^2_{2,3}p^2_{3,1}&\cdots&p^{12}_{1,2}p^{12}_{2,3}p^{12}_{3,1}\\
p^1_{1,2}p^1_{2,3}p^1_{3,2}&p^2_{1,2}p^2_{2,1}p^2_{3,2}&\cdots&p^{12}_{1,2}p^{12}_{2,1}p^{12}_{3,2}-1\\
\end{array}
\right]
\end{align}
\end{tiny}

Then it is easy to verify that $\Phi_{i,j}$ is the rows with each components containing $p^t_{i,j}$ as a factor.

Moreover, a simple calculation shows that Lemma \ref{l3.1} is correct.

\item[(iii)] To verify the effective of ZD-strategies in (\ref{4.4}),
 we assume the strategies for player 1 is
$$
\begin{array}{ccl}
p_{1,1}&=&[0.2,0.3,0.8,0.7,0.5,0.4,0.7,0.9,0.2,\\
~&~&~0.2,0.1,0.9];
\end{array}
$$
the strategies for player 3 is
$$
\begin{array}{ccl}
p_{3,1}&=&[0.15,0.2,0.8,0.85,0.2,0.35,0.7,0.9,0.2,\\
~&~&~0.15,0.55,0.35].
\end{array}
$$

Then the strategy profile dynamics is:
$$
x(t+1)=Lx(t),\quad t\geq 0,
$$
where $L$ is
\begin{tiny}
$$
\begin{array}{ccl}
L&=&\left[
\begin{array}{llllll}
    0.009&0.033&0.128&0.2975&0.04&0.042\\
    0.051&0.132&0.032&0.0525&0.16&0.078\\
    0.018&0.012&0.064&0.0892&0.025&0.07\\
    0.102&0.048&0.016&0.0158&0.1&0.13\\
    0.003&0.015&0.448&0.2082&0.035&0.028\\
    0.017&0.06&0.112&0.0367&0.14&0.052\\
    0.036&0.077&0.032&0.1275&0.04&0.063\\
    0.204&0.308&0.008&0.0225&0.16&0.117\\
    0.072&0.028&0.016&0.0383&0.025&0.105\\
    0.408&0.112&0.004&0.0068&0.1&0.195\\
    0.012&0.035&0.112&0.0893&0.035&0.042\\
    0.068&0.14&0.028&0.0158&0.14&0.078\\
    \end{array}
    \right.\\
~&~&\left.
\begin{array}{llllll}
    0.098&0.1215&0.004&0.0075&0.0055&0.0945\\
    0.042&0.0135&0.016&0.0425&0.0045&0.1755\\
    0.2695&0.2025&0.028&0.0105&0.0083&0.126\\
    0.1155&0.0225&0.112&0.0595&0.0067&0.234\\
    0.1225&0.486&0.008&0.012&0.0413&0.0945\\
    0.0525&0.054&0.032&0.068&0.0338&0.1755\\
    0.042&0.0135&0.016&0.03&0.0495&0.0105\\
    0.018&0.0015&0.064&0.17&0.0405&0.0195\\
    0.1155&0.0225&0.112&0.0420&0.0743&0.014\\
    0.0495&0.0025&0.448&0.238&0.0607&0.0260\\
    0.0525&0.054&0.032&0.048&0.3713&0.0105\\
    0.0225&0.006&0.128&0.272&0.3037&0.0195\\
    \end{array}
    \right].
    \end{array}
$$
\end{tiny}

It is easy to verify that
$$
\rank(L-I_{12})=11.
$$
Moreover, we also have that
$$
\lim_{t\ra \infty} L^t=u\J^T_{12},
$$
where
$$
\begin{array}{ccl}
u&=&[0.0731,0.075,0.0715,0.082,0.126,0.0775,\\
~&~&0.0434,0.1002,0.0475, 0.1278,0.0683,0.1077]^T,
\end{array}
$$
which is the normalized eigenvector of  $L$ with respect to its (unique) eigenvalue $1$.
\end{itemize}
\end{rem}

\section{Application to Networked Evolutionary Games}

This section considers how to design ZD strategies for a player, $i$, in a networked evolutionary game (NEG). We propose a method, called a fictitious opponent player (FOP) conversion.

\subsection{Fictitious Opponent Player}

\begin{dfn}\label{d5.1.1}\cite{che15} An NEG is a triple $((N,E), G, \Pi)$, where $(N,E)$ is a network graph with $N$ as the set of players; $G\in {\cal G}_{[2;k,k]}$ is a symmetric game with two players, called the fundamental network game; $\Pi$ is the strategy updating rule, which describes how each player to update his strategies using his neighborhood information.
\end{dfn}

\begin{rem}\label{r5.1.2}
\begin{itemize}
\item[(i)] $G\in {\cal G}_{[2;k,k]}$ is symmetric, if $S_1=S_2:=S_0$ and for any $x,y\in S_0$
$$
c_1(x,y)=c_2(y,x).
$$
\item[(ii)] If $(i,j)\in E$, then players $i$ and $j$ will play game $G$ at each time. (In this paper only the fixed graph is considered.)
Since $G$ is symmetric, the  the order of two players does not affect the result.

\item[(iii)] Such an NEG is denoted by $G^{ne}=((N,E), G, \Pi)$.
\end{itemize}
\end{rem}

Let player $i\in N$, and  $\deg(i)=d$. Then he may consider $N\backslash\{i\}$ as one player, called the FOP of $i$, denoted by $p_{-i}$. Assume $|S_0|=k$, the neighbors' strategies can be considered as the strategies of $p_{-i}$. That is, $p_{-i}$ has totally $k^d$ strategies.

In fact, we need not to distinct different neighbors, hence if $S_0=\{s_1,s_2, \cdots,s_k\}$, then the set of strategies of $p_{-i}$, denoted by $S_{-i}$, is
\begin{align}\label{5.1.1}
S_{-i}=\{\underbrace{s_1s_1\cdots s_1}_d, \underbrace{s_1s_1\cdots s_2}_d,\cdots,\underbrace{s_ks_k\cdots s_k}_d\}.
\end{align}

Each $s_*\in S_{-i}$ can be expressed as
$$
s_*=(\underbrace{s_1s_1\cdots s_1}_{d_1}\underbrace{s_2s_2\cdots s_2}_{d_2}\cdots\underbrace{s_ks_k\cdots s_k}_{d_k}),
$$
where $d_1+d_2+\cdots+d_k=d$.
Hence, we can also express $s_*$ by $(d_1,d_2,\cdots,d_k)$, which means $s_j$ has been used by $d_j$ neighbors, $1\leq j\leq k$. Using this notation, we have
\begin{align}\label{5.1.2}
S_{-i}=\{(d_1,d_2,\cdots,d_k)\;|\; d_j\geq 0,\;\forall j;\;\dsum_{j=1}^kd_j=d\}.
\end{align}

It is easy to verify that define the strategies of $p_{-i}$ in this way, by ignoring the order of neighbors, the total number of strategies is reduced from $k^d$ to
$$
\frac{(k+d-1)!}{(k-1)!d!}.
$$
This treatment reduces the computational complexity.

From the point of view of player $i$, the NEG is equivalent to a game between him and $p_{-i}$, who has the set of strategies $S_{-i}$ defined by (\ref{5.1.2}).
Let $s_*=(d_1,d_2,\cdots,d_k)\in S_{-i}$. Then the payoff functions for $c_i$ and player $p_{-i}$', denoted by $c_{-i}$, are
\begin{align}\label{5.1.3}
\begin{array}{l}
c_i(x_i,s_*)=\dsum_{j=1}^kd_j(c_i(x_i,s_j),\\
c_{-i}(x_i,s_*)=\dsum_{j=1}^kd_j(c_j(x_i,s_j).\\
\end{array}
\end{align}

Note that the FOP formulation is particularly suitable for using ZD strategies, because it is not effected by the structure and size of network graph, even though the size might be $\infty$. As long as the stationary distribution of the overall network exists, ZD strategies are still applicable. Moreover, it is easily designable.

\subsection{ZD Strategies for NEGs}

This section considers how to design ZD-strategies for NEGs. We describe the process through an example.

Consider prisoner's dilemma, $G$. The two strategies for both players are cooperation ($C$) and defect ($D$). Their payoffs are described in Table \ref{Tab.5.2.1}, where, as a convention, $T>R>P>S$.

\vskip 5mm

\begin{table}
\centering \caption{Payoff bi-matrix}\label{Tab.5.2.1} 
\begin{tabular}{|c||c|c|}
\hline $P_1\backslash P_2$&$C$&$D$\\
\hline
\hline $C$&$R,~R$&S,~T\\
\hline $D$&$T,~S$&$P,P$\\
\hline
\end{tabular}
\end{table}

\vskip 5mm

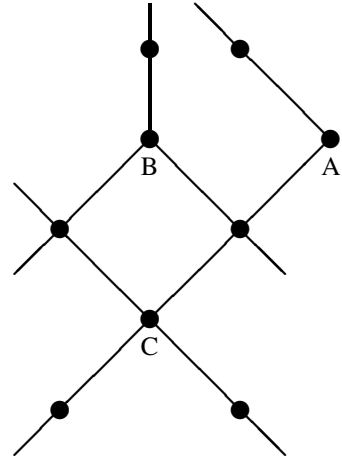
\begin{figure}
\centering
\setlength{\unitlength}{6mm}
\begin{picture}(8,10)\thicklines
\put(0,0){\line(1,1){7}}
\put(0,6){\line(1,-1){6}}
\put(3,7){\line(-1,-1){3}}
\put(3,7){\line(1,-1){3}}
\put(3,7){\line(0,1){3}}
\put(7,7){\line(-1,1){3}}
\put(1,1){\circle*{0.4}}
\put(5,1){\circle*{0.4}}
\put(3,3){\circle*{0.4}}
\put(1,5){\circle*{0.4}}
\put(5,5){\circle*{0.4}}
\put(3,7){\circle*{0.4}}
\put(7,7){\circle*{0.4}}
\put(5,9){\circle*{0.4}}
\put(3,9){\circle*{0.4}}
\put(2.8,6.2){B}
\put(6.8,6.2){A}
\put(2.8,2.2){C}
\end{picture}
\caption{Networked Prisoners' Dilemma}\label{Fig.5.2.1}
\end{figure}

Now consider a networked evolutionary prisoners' dilemma, denoted by $G^{ne}$. The network graph, depicted by Fig. \ref{Fig.5.2.1}, is non-homogeneous.

\vskip 2mm

\vskip 2mm

\begin{enumerate}

\item Consider player A. Since $\deg(A)=2$,
The set of strategies of $p_{-A}$ is
$$
S_{-A}=\{(CC),(CD),(DD)\};
$$
Using (\ref{5.1.3}), the payoff vectors for $c_A$ and $c_{-A}$ are, respectively,
$$
\begin{array}{l}
V^c_A=(2R,R+S,2S,2T,T+P,2P),\\
V^c_{-A}=(2R,R+T,2T,2S,S+P,2P).
\end{array}
$$
It is easy to calculate that $\kappa=6$, and
$$
\Phi_{1,1}=\{1,2,3\},
$$
$$
\chi_{1,1}=(1,1,1,0,0,0).
$$
\begin{itemize}
\item Pinning Strategy:
To get $Ec_{-A}=r$, the ZD strategy of player A can be designed as
$$
(p^1_{1,1},p^2_{1,1},\cdots,p^{6}_{1,1})=
\mu(V^c_{-A}-r\J_{6}^T)-\chi_{11}.
$$
\item Extortion Strategy:
To get $Ec_A-r=\kappa (Ec_{-A}-r)$ with $\kappa >1$, the ZD strategy of player A can be designed as
$$
\begin{array}{l}
(p^1_{1,1},p^2_{1,1},\cdots,p^{6}_{1,1})\\
=\mu\left( (V^c_{A}-r\J_{6}^T)- \kappa (V^c_{-A}-r\J_{6}^T)\right) -\chi_{11}.
\end{array}
$$
\end{itemize}

\item Consider player B. Since $\deg(B)=3$,
The set of strategies of $p_{-B}$ is
$$
S_{-B}=\{(CCC),(CCD),(CDD),(DDD)\};
$$
Using (\ref{5.1.3}), the payoff vectors for $c_B$ and $c_{-B}$ are, respectively,
$$
\begin{array}{ccl}
V^c_B&=&(3R,2R+S,R+2S,3S,3T,\\
~&~&~2T+P,T+2P,3P),\\
V^c_{-B}&=&(3R,2R+T,R+2T,3T,\\
~&~&~3S,2S+P,S+2P,3P).
\end{array}
$$

We also have $\kappa=10$ and
$$
\Phi_{1,1}=\{1,2,3,4,5\},
$$
$$
\chi_{1,1}=(1,1,1,1,1,0,0,0,0,0).
$$

The design of ZD strategies is similar to the one for $A$.

%
%
%
%
%
%
%
%

\item Consider player C. Since $\deg(C)=4$,
The set of strategies of $p_{-C}$ is
$$
\begin{array}{ccl}
S_{-C}&=&\{(CCCC),(CCCD),(CCDD),\\
~&~&~(CDDD),(DDDD)\};
\end{array}
$$
Using (\ref{5.1.3}), the payoff vectors for $c_C$ and $c_{-C}$ are, respectively,

$$
\begin{array}{ccl}
V^c_C&=&(4R, 3R+S,2R+2S,R+3S,4S,4T,\\
~&~&~3T+P,2T+2P,T+3P, 4P),\\
V^c_{-C}&=&(4R, 3R+T, 2R+2T,R+3T,4T,4S,\\
~&~&~3S+P,2S+2P,S+3P,4P).
\end{array}
$$

It is easy to calculate that $\kappa=12$ and
$$
\Phi_{1,1}=\{1,2,3,4,5,6\},
$$
$$
\chi_{1,1}=(1,1,1,1,1,1,0,0,0,0,0,0).
$$

The design of ZD strategies is similar to the one for $A$, or $B$.
\end{enumerate}

\section{Conclusion}

This paper extends the technique of ZD strategies proposed by  Press and Duson to general finite games. Using STP, a fundamental formula is presented to design ZD strategies for general finite games. In addition to the generality, it causes the design procedure very easy. Condition for the rationality of ZD strategies is presented. Then, necessary and sufficient condition for the effectiveness of the designed ZD strategies is also obtained, which put the ZD technique on a solid foundation. Some numerical examples are presented to demonstrate the efficiency of the method proposed in this paper.

Finally, the NEGs are considered. A new concept, called FOP, is proposed as the opponent player for a preassigned player $i$. Using it, the ZD strategies for player $i$ is designed for the game between himself and his FOP. It is a surprising fact that one single player may be able to challenge whole network by using ZD strategies, no matter how large the network is.

Some important issues remained for further study, such as:
\begin{itemize}
\item[(i)] How to design  rational ZD strategies for a complicated game?
\item[(ii)] How to predict the efficiency of ZD strategies for a player?
\item[(iii)] What will happen when ZD strategies are used to against ZD strategies?
\end{itemize}

\section*{Appendix}

\begin{enumerate}
\item

\begin{IEEEproof} of Proposition \ref{p2.3.3}:
\begin{itemize}
\item[(i)] Since $L$ has unique eigenvalue $1$, $\rank(M)=\kappa-1$. Hence there exists at least one $(\kappa-1)\times (\kappa-1)$ miner of $M$, which is nonsingular. Hence, $M^*\neq 0$. Observing that
\begin{align}\label{A.1.1}
MM^*=\det(M)=0,
\end{align}
and $\rank(M)=\kappa-1$, it follows that $\rank(M^*)=1$.

\item[(ii)]  Assume there exists $1\leq j\leq n$ such that $\Col_j(M^*)=0$. Consider $M\backslash \{\Row_j(M)\}$, which is obtained from $M$ by deleting its $j$ th row. Then all its $(\kappa-1)\times (\kappa-1)$ minors have zero determinants. That is, $M\backslash \{\Row_j(M)\}$ is row-dependent.

    To get contradiction, we show that any $\kappa-1$ rows of $M$ are linearly independent. Since
    $\dsum_{i=1}^{\kappa}\Row_i(M)={\bf 0}^T_{\kappa}$, $\Row_j(M)=-\dsum_{i\neq j}\Row_i(M)$. If $\rank(M\backslash \{\Row_j(M)\})<\kappa-1$, then $\rank(M)<\kappa-1$, which is a contradiction.
\end{itemize}
\end{IEEEproof}

\item

\begin{IEEEproof} of Proposition \ref{p2.3.4}:

First we show that the limit exists. Since $\{x(t)\;|\; t=1,2,\cdots\}\subset \Upsilon^{\kappa}$ and  $\Upsilon^{\kappa}$ is a compact set, there exists a subsequence $\{x_{t_i}\;|\;i=1,2,\cdots\}$ such that
$$
\lim_{i\ra \infty}x_{t_i}=x^*\in \Upsilon^{\kappa}.
$$
Note that $\lim_{t\ra\infty}L^t=P$, denote by $x^0=Px^*$, we claim that
\begin{align}\label{A.2.1}
\lim_{t\ra \infty}x_{t}=x^0.
\end{align}
Given any $\epsilon>0$, there exists $N_1$ such that when $t_i>N_1$
$$
\|x_{t_i}-x^*\|<\sqrt{\epsilon};
$$
and there exists $N_2>0$ such that when $t>N_2$
$$
\|M^t-P\|<\sqrt{\epsilon}.
$$
Choose  an element $t_{i_0}>N_1$ from the subsequence and set
$$
N_3=t_{i_0}>N_1.
$$
Assume  $t>N_2+N_3$, then
$$
x(t)=M^{t-N_3}x(t_{i_0}).
$$
Since $t-N_3>N_2$ and $N_3=t_{i_0}>N_1$, it follows that
$$
\|x(t)-x^0\|<(\sqrt{\epsilon})^2=\epsilon.
$$
(\ref{A.2.1}) follows. It is also clear that $x^0=x^*$. Moreover, $Lx^*=x^*$ and $Px^*=x^*$. Now since $P=uv^T\in \Upsilon_{\kappa\times \kappa}$, without loss of generality, we can normalize $u$ by replacing $u$ by $u/\|u\|$. Then $v=\J_{\kappa}$. Moreover,
$$
x*=Px^*=uv^Tx^*=u.
$$

\end{IEEEproof}

\item

\begin{IEEEproof} of Proposition \ref{p2.3.5}:

Note that $MM^*=\det(M)=0$, and $Mu-0$. Since $\rank(M)=\kappa-1$,
the solutions of equation $Mx=0$ is a one-dimensional subspace. Now each column of $M^*$ is a solution, the conclusion follows.

\end{IEEEproof}

\begin{IEEEproof} of Lemma \ref{l3.1}:

Recall the PEE (\ref{2.2.7}), using (\ref{2.2.701}), a straightforward computation shows that

\begin{align}\label{A4.1}
\begin{array}{ccl}
M&=&L-I_{\kappa}\\
~&=&[
p^1-\d_{\kappa}^1,p^2-\d_{\kappa}^2,\cdots,p^{\kappa}-\d_{\kappa}^{\kappa}],
\end{array}
\end{align}
where
$$
p^r=\begin{array}{lllll}
p^r_{1,1}&p^r_{2,1}&\cdots&p^r_{n-1,1}&p^r_{n,1}\\
p^r_{1,1}&p^r_{2,1}&\cdots&p^r_{n-1,1}&p^r_{n,2}\\
\vdots&~&~&~&~\\
p^r_{1,1}&p^r_{2,1}&\cdots&p^r_{n-1,1}&p^r_{n,k_n}\\
p^r_{1,1}&p^r_{2,1}&\cdots&p^r_{n-1,2}&p^r_{n,1}\\
\vdots&~&~&~&~\\
p^r_{1,1}&p^r_{2,1}&\cdots&p^r_{n-1,k_{n-1}}&p^r_{n,k_n}\\
\vdots&~&~&~&~\\
p^r_{1,k_1}&p^r_{2,k_2}&\cdots&p^r_{n-1,k_{n-1}}&p^r_{n,k_n}\\
\end{array}
$$

It is easy to figure out that the factors $\{p^r_{i,1},p^r_{i,2},\cdots,p^r_{i,k_i}\}$ appear in $p^r$  as
\begin{align}\label{A4.2}
p^r_i=\left(\J_{\kappa_{(i)}}\right)
\begin{bmatrix}
p^r_{i,1}\\
p^r_{i,2}\\
\vdots\\
p^r_{i,k_i}
\end{bmatrix}\left(\J_{\kappa^{(i)}}\right).
\end{align}

It can be seen by a careful observation that
\begin{itemize}
\item[(i)] $\Phi_{i,j}$ is the rows where $p^r_{i,j}$ appears.
\item[(ii)] Let $s\in \Phi_{i,j}$. Denote
$$
\Row_s(M)=\left(c^1_{i,j}(s)p^1_{i,j}, c^2_{i,j}(s)p^2_{i,j},\cdots,c^{\kappa}_{i,j}(s)p^{\kappa}_{i,j}\right),
$$
where $c^{\ell}_{i,j}(s)$ is used to express the coefficient of $p^{\ell}_{i,j}$, which is the product of some
$p^{\ell}_{p,q}$, $(p,q)\neq (i,j)$. Then
$$
\dsum_{s\in \Phi_{i,j}}c^{\ell}_{i,j}(s)=1,\quad \ell=1,2,\cdots,\kappa.
$$
\item[(iii)] If $s\in \Phi_{i,j}$, the in the row summation $\Xi$ a constant $-[\d^s_{\kappa}]^T$ is added. Adding all such constants together yields $\xi_{i,j}$.
\end{itemize}
The conclusion follows from (i), (ii), and (iii).
\end{IEEEproof}

\item

\begin{IEEEproof} of  Theorem \ref{t3.10}:

(Necessity) It is obvious that to make ZD strategies workable the existence of stationary distribution is necessary. That is,
\begin{align}\label{A5.1}
\lim_{t\ra \infty}L^tx_0=u,\quad \forall x_0\in \Upsilon^{\kappa}.
\end{align}
We first prove $\lim_{t\ra \infty}L^t$ exists. Since $\Upsilon_{\kappa\times \kappa}$ is a compact set, if
$\lim_{t\ra \infty}L^t$ does not exist, there must be at least two subsequences $\{L^{n_i}\}$, and $\{L^{m_i}\}$, such that
$$
\begin{array}{l}
\lim_{i\ra \infty}L^{n_i}=P_1,\\
\lim_{i\ra \infty}L^{m_i}=P_2,\\
\end{array}
$$
and $P_1\neq P_2$. Say, $\Col_s(P_1)\neq \Col_s(P_2)$, choosing $x_0=\d_{\kappa}^s$, then it violates (\ref{A5.1}).

Hence we have
$$
\lim_{t\ra\infty}L^t=P.
$$
Again, because of (\ref{A5.1}) $P$ should have the form that
$P=[u,u,\cdots,u]$, the conclusion is obvious.

As for the condition (ii), if $\rank(L-I_{\kappa})<\kappa-1$, then $M^*=0$ is a zero matrix.
Then (\ref{3.7}) fails. Hence (\ref{3.7}) can never be obtained from (\ref{3.8}), and ZD strategies do not work.

(Sufficiency) Replacing any row $s\in \Phi_{i,j}$ of matrix $M=L-I_{\kappa}$ by $\Xi_{i,j}$, then condition (ii) ensures (\ref{2.3.7}). Using (\ref{3.8}) and expanding the determinant via replaced row, (\ref{3.7}) follows.
\end{IEEEproof}
\end{enumerate}

\begin{thebibliography}{00}
%
\bibitem{cao19} M. Cao, C. Tang, Y. Liu, F. Lin, Z. Chen, Application of ZD strategy in mining pool game, {\it Proc. 38th CCC}, 880-885, 2019.
%
\bibitem{che11} D. Cheng, H. Qi, Z. Li, {\it Analysis and Control of Boolean Networks: A Semi-tensor Product Approach}, Springer, London, 2011.
%
\bibitem{che12} D. Cheng, H. Qi, Y. Zhao, {\it An Introduction to Semi-tensor Product of Matrices and Its Applications}, World Scientific, Singapore, 2012.
%
\bibitem{che14} D. Cheng, On finite potential games, {\it Automatica}, Vol. 50, No. 7, 1793-1801, 2014.
%
\bibitem{che15} D. Cheng, F. He, H. Qi, T. Xu, Modeling, analysis and control of networked evolutionary games,
{\it IEEE Trans. Aut. Contr.}, Vol. 60, No. 9, 2402-2415, 2015.
%
\bibitem{che16}  D. Cheng, T. Liu, K. Zhang, On decomposed subspaces of finite games, {\it IEEE Trans. Aut. Contr.}, Vol. 61, No. 11, 3651-3656, 2016.
%
\bibitem{che17}       D. Cheng, T. Liu, Linear representation of symmetric games, {\it IET Contr. Theory Appl.}, Vol. 11, No. 18, 3278-3287, 2017.
%
\bibitem{chepr} D. Cheng, Y. Wu, G. Zhao, S. Fu, A Comprehensive Survey on STP Approach to Finite Games,
{\it Sys. Sci. Compl.}, http://arxiv.org/abs/2106.16086, (submitted).

\bibitem{hao18}    Y. Hao, D. Cheng, On skew-symmetric games, {\it Journal of the Franklin Institute}, Vol. 355, 3196-3220, 2018.
%
\bibitem{hao14} D. Hao, Z. Rong, T. Zhou, Zero-determinant strategy: An underway revolution in game theory, {\it Chin. Phys. B}, Vol. 23, No. 7, 078905, 2014.
%
\bibitem{hor86} R.A. Horn, C.R. Johnson, {\it Matrix Analysis}, Cambridge, Campbidge Univ. Press, 1986.
%
\bibitem{li18} H. Li, X. Ding, Q. Yang, Y. Zhou, 2018. Algebraic formulation and Nash equilibrium of competitive diffusion games, {\it Dyn. Game Appl.}, Vol. 8, 423-433, 2018.
%
\bibitem{mca16} A. Mcvoy, C. Hauert, Auticratic strategies for iterated games with arbitrary action spaces, {\it Proc. Natl. Acad. Sci.}, Vol. 113, doi:10,1073/pnas.1520163113, 2016.

\bibitem{pre12} W. Press, F.J. Duson, Iterated prosoner's dilemma containes strategies that dominate any evolutionary opponent, {\it Proc. Natl. Acad. Sci.}, Vol. 109, No. 26, 10409-10413, 2012.
%
\bibitem{he16} X. He, H. Dai, P. Ning, R. Dutta, Zero-determinant strategies for multi-player multi-action iterated games. {\it IEEE Signal Proc. Lett.}, Vol. 23, No. 3, 311-315, 2016.
%
\bibitem{hil14} C. Hilbe, B. Wu, A. Traulsen,A. M.A. Nowak, Cooperation and control in multplayer social dilemmas,
{\it Proc. Natl. Acad. Sci.}, Vol. 111, No. 46, 16425-16430, 2014.
%
\bibitem{pan15} L. Pan, D. Hao, Z. Rong, T. Zhou, Zero-determinant strategies in iterated public goods game, {\it Sci. Rep.}, Vol. 5, 13096, 2015.
%
\bibitem{szo14} A. Szolnoki, M. Perc, Evolution of extortion in structured populations, {\it Phys. Rev. E}, Vol. 89, No. 2, 022804, 2014.
%
\bibitem{tan21} R. Tan, Q. Su, B. Wu, L. Wang, Payoff control in repeated games, {\it Proc. 33rd Chinese CDC}, 997-1005, 2021.
%
\bibitem{wan17} J. Wang, J. Guo, H. liu, A. Shen, Evolution of zero-determinant strategy on iterated snowdrift game, {\it Acta Phus. Sin.}, Vol. 66, No. 18, 180203, 2017.
%
\bibitem{zha21} J. Zhang, J. Lou, J. Qiu, J. Lu, Dynamics and covergence of huper-networked evolutionary games with time delay is strategies, {Information Sciences}, Vol. 563, 166-182, 2021.
%
\end{thebibliography}
\end{document}